\documentclass[aps,prl,twocolumn, showpacs,superscriptaddress,citeauthorscript]{revtex4-2} 

\usepackage{graphicx}  

\usepackage{amssymb}   
\usepackage{amsmath}
\usepackage{gensymb}

\usepackage{hyperref}
\usepackage[T1]{fontenc}

\begin{document}

\title{Real-time milli-Kelvin thermometry in a semiconductor qubit architecture}

\author{V. Champain}
\email{E-mail: victor.champain@cea.fr}
\author{V. Schmitt}
\affiliation{Univ. Grenoble Alpes, CEA, Grenoble INP, IRIG-Pheliqs, Grenoble, France.}
\author{B. Bertrand}
\author{H. Niebojewski}
\affiliation{Univ. Grenoble Alpes, CEA, LETI, Minatec Campus, Grenoble, France.}
\author{R. Maurand}
\author{X. Jehl}
\author{C. B. Winkelmann}
\author{S. De Franceschi}
\author{B. Brun}
\email{E-mail: boris.brun-barriere@cea.fr}
\affiliation{Univ. Grenoble Alpes, CEA, Grenoble INP, IRIG-Pheliqs, Grenoble, France.}

\date{\today}

\begin{abstract}

We report local time-resolved thermometry in a silicon nanowire quantum dot device designed to host a linear array of spin qubits.  Using two alternative measurement schemes based on rf reflectometry, we are able to probe either local electron or bosonic bath temperatures with $\mu$s-scale time resolution and a noise equivalent temperature of $3$ $\rm mK/\sqrt{\rm Hz}$. Following the application of short microwave pulses, causing local periodic heating, time-dependent thermometry can track the dynamics of thermal excitation and relaxation, revealing clearly different characteristic time scales. This work opens important prospects to investigate the out-of-equilibrium thermal properties of semiconductor quantum electronic devices operating at very low temperature. In particular, it may provide a powerful handle to understand heating effects recently observed in semiconductor spin-qubit systems.   
\end{abstract}
\maketitle
\section{Introduction}

Quantum electronics embraces a large variety of devices whose functionality relies on quantum mechanical properties such as size quantization, phase coherence, entanglement, etc. Cryogenic operation is generally required for the emergence of such properties and, at low temperature,  the time scales for energy exchange and thermalization processes tend to increase dramatically, leading to metastable regimes and to the co-existence of thermal (and non-thermal) baths strongly out of equilibrium. The understanding and control of these thermodynamical aspects is not only mandatory for the development and operation of bolometers, cryogenic thermometers or coolers; it can also be particularly crucial for improving the performance and scalability of solid-state devices for quantum sensing and quantum computing. To this aim, access to thermometers capable of measuring local temperatures faster than the characteristic time scales of heat exchange dynamics is of primary importance\cite{noah_cmos_2024}.  

In this work, we address this point in the context of semiconductor quantum-dot (QD) devices, largely motivated by their prospects for scalable spin-based quantum computing. The operation of semiconductor spin qubits requires applying high-frequency signals that unavoidably dissipate energy into the qubit environment due to the Joule effect and dielectric losses. 
This raises severe practical issues regarding spin qubits addressability, since a significant temperature dependence of their Larmor frequencies has recently been reported \cite{undseth_hotter_2023}.
Fast temperature changes due to unintentional heating can lead to spin decay and dephasing \cite{lawrie_simultaneous_2023,takeda_optimized_2018}, thereby reducing gate and readout fidelities \cite{kawakami_gate_2016}. This problem worsens for increasing numbers of qubits and corresponding control gates, leading to a clear bottleneck for large-scale integration. A higher operation temperature can mitigate the impact of local heating \cite{undseth_hotter_2023} but at the price of a reduced spin coherence time \cite{yang_operation_2020,camenzind_hole_2022}.  Therefore, understanding the modality and dynamics of heating from microwave excitation is a necessary step to devise more efficient measures to preserve qubit performance.  To this end, access to fast time-domain thermometry compatible with mK temperature and hundreds of mT fields can be a clear asset.

The most advanced local thermometry techniques in quantum circuits rely on metal-superconductor junctions \cite{gasparinetti_fast_2015,gumus_calorimetry_2023,giazotto_opportunities_2006}. These thermometers are incompatible with spin-based quantum processors, due to integration constrains and to their susceptibility to magnetic fields. Semiconductor QD thermometers\cite{giazotto_opportunities_2006,beenakker_theory_1991,maradan_gaas_2014,nicoli_quantum_2019,torresani_nongalvanic_2013}, on the other hand, do not suffer from these limitations. Moreover, they can measure the temperature of an electronic reservoir without requiring galvanic coupling \cite{ torresani_nongalvanic_2013, ahmed_primary_2018,chawner_nongalvanic_2021}. 

Here we apply this approach to silicon QD devices. By adjusting the tunnel coupling between the QD thermometer and a probed electronic reservoir, we are able to preserve its sensitivity down to the base temperature (55 mK) of a dilution refrigerator. In addition, by adopting an isolated double-dot configuration of the QD thermometer, we extend our thermometry capabilities to probe the local bosonic temperature. We simultaneously show that this thermometer can perform time-domain measurements at the microsecond scale.

\section{Device and operation modes}
The QD devices consist of silicon-on-insulator metal-oxide-semiconductor (MOS) nanowires with a rectangular cross-section of 80 $\times$16 nm$^2$. The nanowires are covered by a set of  eight parallel gates with a pitch of 80 nm, see example in Fig. 1(a). Thanks to p-doped leads, holes are accumulated under negatively-biased gates as schematically shown in Fig. 1(b).

\begin{figure}
    \centering
    \includegraphics[width=\linewidth]{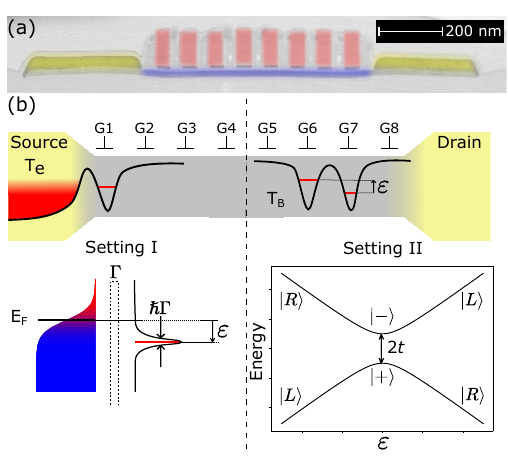}\label{fig:fig1}
    \caption{ \textbf{Device and charge configuration for two thermometry techniques.} \textbf{(a)} False-color tunneling electron micrograph of a representative MOS device with a silicon nanowire channel (blue), highly doped conductive leads (yellow) and eight parallel gates (red). Note that the gates can be considered metallic, however the contacts on the leads and gates are made of tungsten vias, that is superconducting at base temperature. Three different samples were measured in this work (see Supp. Mat. for details)\cite{SuppMat}. \textbf{(b)} Schematic of a 8-gate device illustrating the two thermometry methods. Source and drain (yellow) are highly p-doped to form metallic leads acting as hole reservoirs. Quantum dots (QDs) can be defined along the undoped silicon channel  by means of gates G1 to G8. In the first type of thermometer (setting I), G1 defines a single QD tunnel-coupled to the source reservoir as illustrated by the hole-energy diagram just below. The hole reservoir is modelled by a Fermi distribution with electronic temperature $T_e$, while the dot is modelled by a single level whose linewidth is determined by the tunnel rate $\Gamma$. The detuning, $\varepsilon$, is defined here as the difference between the QD electrochemical potential and the Fermi level in the reservoir.  In the second thermometer (setting II), a double QD isolated from the reservoirs is defined under G6 and G7. Both QDs are modelled by single levels with a tunnel coupling $t$ and detuning $\varepsilon$. The corresponding energy diagram is shown just below, with $|L\rangle$ ($|R\rangle$) representing the state localized in the left (right) QD at large detuning, and $|+\rangle $ ($|-\rangle$) the hybridized bonding (antibonding) state at zero detuning. The coupling to the leads is completely suppressed by means of the two side gates G5 and G8. }   
\end{figure}

We investigate two different device settings. In the first setting (I), we define a single QD tunnel-coupled to the hole reservoir in one the two degenerately-doped contact leads, as shown on the left hand side of \hyperref[fig:fig1]{Fig. 1(b)}. The impedance of a lumped element LC circuit connected to the same lead is sensitive to charge tunneling between the QD and the lead. Both dissipative and capacitive changes can be detected through reflectometry measurements \cite{vigneau_probing_2023,gonzalez-zalba_probing_2015,house_radio_2015}.

The second setting (II) requires biasing four consecutive gates, as shown on the right hand side of \hyperref[fig:fig1]{Fig. 1(b)}. The two inner gates (G6, G7) define a double QD, whereas the outer ones (G5, G8) allow fully isolating this double QD from the remaining part of the channel and from the leads\cite{bertrand_quantum_2015,eenink_tunable_2019}.  Read-out is now performed by reflectometry on an inner gate electrode (G7), with interdot tunneling resulting in a quantum capacitance contribution \cite{vigneau_probing_2023,gonzalez-zalba_probing_2015,colless_dispersive_2013}.

The tank circuits used for RF-reflectometry are formed by classical surface-mount inductors and parasitic capacitances. While different values of inductance have been used, all resonances are around $400$ MHz, with few-to-several MHz bandwidth (see Supp. Mat. for details)\cite{SuppMat}. The reflectometry measurements are performed in a weak coupling, dispersive regime, in which the phase shift of the reflected signal is directly proportional to the frequency shift, and thus to the quantum capacitance \cite{blais_circuit_2021,petersson_charge_2010,vigneau_probing_2023,colless_dispersive_2013}.

\section{Quantum-dot thermometry: basic principles and experimental implementation}
\label{sec:sec2}

In both experimental configurations, the quantum capacitance contribution is expected to depend on temperature. 
In setting I, where a single level is tunnel-coupled to a Fermi sea, the quantum capacitance reads \cite{chawner_nongalvanic_2021,torresani_nongalvanic_2013,beenakker_theory_1991}: (see Supp. Mat. for details)\cite{SuppMat}
\begin{align}
    C_q(\varepsilon) = \alpha^2 e^2\int_{-\infty}^{+\infty } & \left (  \frac{1}{4k_BT_e} \cosh\left (\frac{e \alpha \varepsilon -E}{2k_BT_e}\right )^{-2}\right ) \notag \\
    \times &\left ( \frac{\hbar \Gamma}{\hbar^2\Gamma^2 + E^2}\right )\text{d}E.
    \label{eq:fermi}
\end{align}
Here $e$ is the electron charge, $k_B$ the Boltzmann constant, $\hbar$ the reduced Planck constant,  $\alpha$ the gate lever-arm parameter, $\varepsilon$ the detuning (in gate voltage), and $\Gamma$ the dot-lead tunnel rate. The relevant temperature is the electronic temperature of the lead, $T_e$. The quantum capacitance is therefore a convolution of two peaks: the derivative of the Fermi distribution, broadened by $k_B T_e$, and the dot density of states, modeled by a Lorentzian of width $\hbar\Gamma$. Thermal broadening dominates for $\hbar \Gamma < k_B T_e$. 

In setting II, hybridization of the orbitals in the left and right QDs results in a two-level system in the canonical ensemble. The corresponding energy diagram as a function of the gate-dependent level detuning, $\varepsilon$, is characterized by an avoided level crossing due to the inter-dot tunnel coupling $t$. The quantum capacitances associated to the bonding and anti-bonding states have opposite signs since they are 
directly related to the curvatures of the corresponding energy-vs-detuning relations  \cite{petersson_charge_2010}. The total capacitance is given by the sum of the two quantum capacitances weighted by the population of the bonding and anti-bonding states $P_{|+\rangle}$, $P_{|-\rangle}$ (see Supp. Mat. for details)\cite{SuppMat}, i.e.:  \cite{petersson_charge_2010,van_der_vaart_resonant_1995,beenakker_theory_1991} 
\begin{equation}
    C_q(\varepsilon) = \alpha^2 e^2\frac{2t^2}{\left ( (\alpha e\varepsilon)^2+4t^2\right )^{3/2}}\tanh\left(  \frac{\left ( (\alpha e\varepsilon)^2+4t^2\right )^{1/2}}{2k_B T_{B}}\right).
    \label{eq:maxboltz}
\end{equation}
Where $T_B$ stands for the Boltzmann temperature of the double dot, given by the relative population in both states $k_BT_B = 2t/\ln{(P_{|+\rangle}/P_{|-\rangle}})$. At thermal equilibrium with the thermal bath, i.e. in the absence of external charge drive, the Boltzmann temperature equals the bath temperature. We notice that $C_q(\varepsilon)$ is peaked at $\varepsilon = 0$ with a linewidth solely determined by the tunnel coupling $t$, and an amplitude that becomes strongly temperature dependent in the limit $t\alt k_B T_{B}$.

\begin{figure}
   \centering    
   \includegraphics[width=\linewidth]{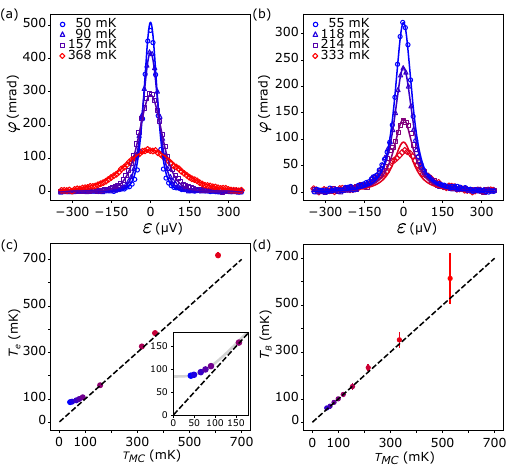}\label{fig:fig2} 
   \caption{\textbf{Experimental demonstration of the two thermometry techniques} \textbf{(a)} Setting I: Phase of the reflectometry signal from the tank circuit as a function of detuning, for a dot-lead transition at different mixing chamber (MC) temperatures. The peak amplitude is decreasing with temperature while the peak width increases.  \textbf{(b)} Setting II: Phase of the reflectometry signal for an interdot transition as a function of MC temperature. The peak amplitude is decreasing, while the width remains unchanged. \textbf{(c)} $T_{MC}$ dependence of the electronic temperature, $T_e$, in the source reservoir as obtained from fitting to the Eq. (\ref{eq:fermi}). \textbf{Inset:} Closeup of the low-temperature saturation. \textbf{(d)}  $T_{MC}$ dependence of  $T_{B}$, as obtained from fitting the the Eq. (\ref{eq:maxboltz}).}
\end{figure}

Both settings are experimentally tested by thermally anchoring the device to the mixing chamber of a dilution refrigerator with a base temperature of 55 mK, and the results are shown in \hyperref[fig:fig2]{Fig 2}. We initially verify that the impedance of the LC-resonator itself is temperature independent in the range of interest. \hyperref[fig:fig2]{Fig. 2(a)} presents the temperature dependence of the reflected signal measured in the case of setting I (left-hand side of \hyperref[fig:fig1]{Fig. 1(b)}). The data are obtained after selecting the dot-lead transition with the narrowest Coulomb peak resonance (see Supp. Mat. for details)\cite{SuppMat}. The phase of the reflected signal displays a peak proportional to the expected peak in $C_q(\varepsilon)$ whose amplitude decreases with the mixing-chamber temperature, $T_{MC}$, while its width increases. This is in very good agreement with Eq. (\ref{eq:fermi}), which we use to fit the experimental peaks (see Supp. Mat. for details)\cite{SuppMat}. We find a tunnel coupling $\Gamma/2\pi = 383\pm38$ MHz, an order of magnitude lower than $k_BT_e/h$ at base temperature, thereby fulfilling the condition for proper temperature sensitivity. 

For the case of setting II (right-hand side of \hyperref[fig:fig1]{Fig. 1(b)}), we first load a few holes under G6 and G7 by applying negative voltages to G6, G7 and G8. Bringing  G8 back to zero voltage results in the trapping a few charges confined to a double QD controlled by G6 and G7. Provided the interdot tunnel coupling is greater than the LC resonator frequency, charge transitions from one dot to the other can be detected by RF reflectometry. In this isolated regime, dot-lead transitions are inhibited and the interdot transitions are the only visible spectroscopic features, extending over the entire stability diagram \cite{eenink_tunable_2019,bertrand_quantum_2015}(see Supp. Mat. for details)\cite{SuppMat}.
As we follow the interdot transition towards more negative voltages, we create deeper electrostatic quantum wells, effectively increasing the barrier height and hence lowering $t$. We can thus achieve a regime where the thermal population of the double-QD two-level system becomes highly sensitive to the bath temperature $T_{B}$. Similarly to the first configuration, we measure the interdot transitions as a function of $T_{MC}$, the results being shown in \hyperref[fig:fig2]{Fig. 2(b)}. Fitting to Eq. \ref{eq:maxboltz} yields a tunnel rate $t/h=2.03\pm0.04$ GHz. Even though the gap due to level repulsion is four times larger than the thermal energy at base temperature, we still have a $2\%$ thermal occupation of the excited state. This ensures a measurable temperature dependence of the quantum capacitance all over the explored temperature range. Yet extending this measurement technique to even lower temperatures would require reducing the tunnel coupling accordingly, as discussed later. Nonetheless, we note that the relatively low tunnel coupling achieved here enables us to operate this non-galvanic thermometer at temperatures an order of magnitude lower than previously reported \cite{chawner_nongalvanic_2021}.

The measured temperatures extracted from the fits are shown in \hyperref[fig:fig2]{Fig. 2(c) and (d)} as a function of $T_{MC}$ in setting I and II, respectively. At relatively high temperature, both $T_e$ and $T_{B}$ follow closely the mixing chamber temperature, except for significant deviations above $500$ mK that can be attributed to the population of higher-energy QD levels, which is not taken into account in our models. 

Below about 100 mK, however, the two experimental settings exhibit different behavior. In setting I, the measured electronic temperature saturates. This trend, commonly observed in cryogenic experiments, is due to the residual electromagnetic noise coming from the circuitry at higher temperature, combined with the vanishing thermal coupling of electrons to their environment \cite{zieve_low-temperature_1998,gasparinetti_probing_2011}. The inset in \hyperref[fig:fig2]{Fig. 2(c)} shows a closeup of these data points, with a fit to a saturation function  $T_e = (T_{\rm MC}^n + T_0^n)^{1/n}$, yielding here $n=3.4$ and a saturation temperature $T_0=84$ mK. In the case of an electron-phonon dominated thermalization \cite{giazotto_opportunities_2006}, one would expect $n=5$ in degenerately doped silicon\cite{zieve_low-temperature_1998}. However, the structure of the lead in our experiment involves also other materials in close vicinity, including metallic/superconducting vias, hindering an accurate thermal modelling.

Interestingly, no such saturation is observed in the case of setting II. Indeed, here the isolated double-QD is not coupled to any Fermi sea and its temperature is determined by the interaction with the local bosonic baths, i.e. lattice phonons and photons in the electromagnetic environment.

\section{Thermometer calibration and optimization}

We now investigate the possibility of performing local thermometry in the time domain. We focus on setting II.  The peak amplitude of the phase shift at zero detuning strongly depends on the bath temperature and it can be measured on a microsecond time scale. This opens the possibility to monitor the time evolution of $T_{B}$, provided thermal equilibration between the QD two-level system and the surrounding bath occurs on a shorter time scale.

\begin{figure}
    \centering
    \includegraphics[width=\linewidth]{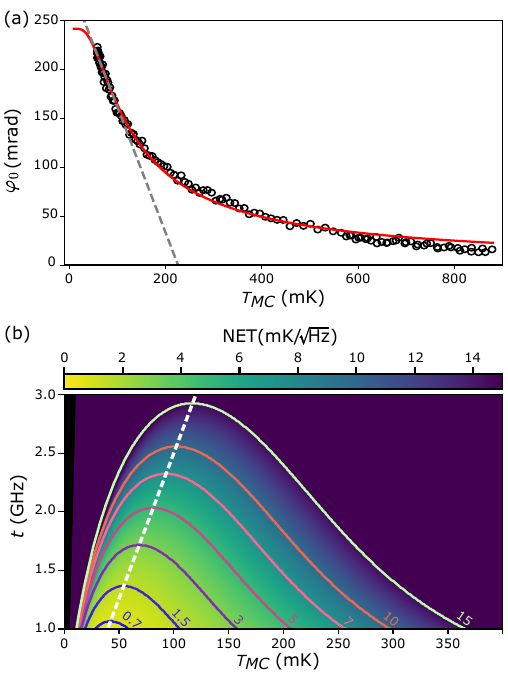}\label{fig:fig3}
    \caption{\textbf{Thermometer calibration and optimal operation regime for setting II.} \textbf{(a)} Peak amplitude, $\varphi_0$, as a function of $T_{MC}$ in the case of setting II. The experimental data (open circles) are fitted to the Eq. (\ref{eq:maxwellzerodetuning}) (red solid line), giving a tunnel rate $t/h = 1.72\pm0.04$ GHz. The grey dashed line is a linear fit to the data in the $55 - 100$ range of highest sensitivity (${\lvert \partial \varphi_0 /\partial T \rvert} = 1.28$ mrad/mK). \textbf{(b)} Calculated noise equivalent temperature, $NET$, versus ($T_{MC}$,$t$). Iso-NET curves are drawn in blue and yellow colors. The white dashed line highlights the $t$-dependence of the NET minimum (i.e., the $T_{MC}$ where $\partial NET /\partial T_{MC} = 0$).}
\end{figure}

To precisely calibrate this thermometer, we measure the reflected signal phase at zero detuning, $\varphi_0$, and vary the mixing-chamber temperature. The resulting calibration curve is shown in \hyperref[fig:fig3]{Fig. 3(a)}. 
From Eq. (\ref{eq:maxboltz}), by taking $\varepsilon = 0$ and, based on \hyperref[fig:fig2]{Fig. 2(d)}, $T_{B} = T_{MC}$,  we obtain 

\begin{align}
    \varphi_0 = \kappa \times \frac{1}{2t}\tanh \left ( \frac{t}{k_B T_{MC}}\right ),\label{eq:maxwellzerodetuning}
\end{align}
where $\kappa$ is a proportionality coefficient that does not depend on temperature.

The data are in good agreement with this model (red solid line in \hyperref[fig:fig3]{Fig. 3(a)}), where  $t$ and $\kappa$ are the free parameters. We note that, in principle, the anti-crossing gap ($2t$) in such a configuration could be independently measured through Landau-Zener-St\"uckelberg interference \cite{stoof_time-dependent_1996,dupont-ferrier_coherent_2013}. Such an additional measurement would make this type of thermometer a primary one, removing the need of a calibration process.

Notably, the model used for the fit in \hyperref[fig:fig3]{Fig. 3(a)} predicts a saturation of the peak amplitude at low temperature, i.e. when the population of the anti-bonding state becomes negligibly small ($\lesssim 0.5\%$). In our experimental conditions, this saturation is not observed because it occurs slightly below the mixing-chamber base temperature. 
\hyperref[fig:fig3]{Fig. 3(a)} also shows that the thermometer sensitivity (proportional to $\lvert {\partial \varphi_0 /\partial T_{MC} }\rvert$) gets maximal close to the fridge base temperature, i.e. just above the expected low-temperature saturation. The dashed grey line in \hyperref[fig:fig3]{Fig. 3(a)} is a linear fit around the point of maximal sensitivity (55-100 mK). 

To further characterize the thermometer performance and deepen our understanding of its optimal operating conditions, we now turn to the evaluation of the noise equivalent temperature ($NET$) and its dependence on $t$ and $T_{MC}$. Since 
$NET =\frac{S_{\varphi\varphi}}{\lvert {\partial \varphi_0 /\partial T_{MC} }\rvert}$, where $S_{\varphi \varphi}$ is the phase noise amplitude, we begin by recording the phase signal at various sampling rates covering a large frequency domain. The measured spectrum reveals a white noise floor at $S_{\varphi \varphi} = 3.9 \pm 0.2\text{ mrad}/\sqrt{\text{Hz}}$, which is consistent with the expected noise from the first amplification stage at 4 K. This gives $NET = 3.0 \pm 0.2\text{ mK}/\sqrt{\text{Hz}}$ in the maximal sensitivity, low-temperature region of \hyperref[fig:fig3]{Fig. 3(a)}. This value is expected to change by varying the interdot tunnel coupling and the temperature range of operation. To fully capture this dependence, implicitly coming from the $\lvert {\partial \varphi_0 /\partial T_{MC} }\rvert$ term, we make use of Eq. (\ref{eq:maxwellzerodetuning}) with the additional hypothesis of a dispersive coupling between the LC resonator and the isolated double QD, which implies\cite{blais_circuit_2021}: 
\begin{equation}
     \kappa = A_0 \left ( \frac{1}{2t + f_r} + \frac{1}{2t - f_r}\right ) \label{eq:dispersivescaling}
 \end{equation}
where $f_r$ is the resonant frequency of the tank circuit and $A_0$ is a constant depending only on circuit parameters , whose value can be deduced from the fit of \hyperref[fig:fig3]{Fig. 3(a)}. 
The resulting $NET(T_{MC}, t)$ map, extrapolated from the data in \hyperref[fig:fig3]{Fig. 3(a)}, is shown in \hyperref[fig:fig3]{Fig. 3(b)}.

For a given tunnel coupling, the thermometer is expected to have an optimal temperature where $NET$ is minimized.  The white dashed line highlights the $t$-dependence of this optimal operation temperature. 
Interestingly, reducing the tunnel coupling at fixed temperature always results in a lower $NET$ no matter what the temperature is. Therefore, operating at weak tunnel coupling is generically beneficial. However, this only holds as long as the dynamics of the LC resonator is slow with respect to interdot tunneling (adiabatic limit), i.e. away from the divergence in Eq. (\ref{eq:dispersivescaling}) for $t \sim 2f_r$. 
For this reason, the calculation in \hyperref[fig:fig3]{Fig. 3(b)} is limited to $t \geq 1$ GHz. (Below this cutoff, other physical mechanisms such as charge noise would also limit the thermometer sensitivity.)
For $t=1$ GHz, the thermometer NET could be as low as $1\text{ mK}/\sqrt{\text{Hz}}$ at $30$ mK. Moreover, a better impedance matching between the device and the LC resonator would result in a larger $A_0$ and hence an even lower NET, thereby approaching and possibly exceeding the performance of state-of-the-art metal-superconductor thermometers \cite{gasparinetti_fast_2015}.

\section{ Proof-of-concept: real-time thermometry}

Finally, we address the use of our device for real-time thermometry. The measured $NET$ is in principle too large to allow stochastic temperature fluctuations to be resolved at a microsecond time scale.  However, the developed thermometry could still be applied to probe the dynamics of local thermal exchange caused by deterministic and periodic heating events, such as those associated with the sequences of microwave bursts typically employed in the operation of spin qubit devices. Under these conditions, the signal to noise ratio can be increased by averaging over multiple cycles. As a result, the thermometry bandwidth is no longer limited by the $NET$ but by the response time of the system itself.

We identify two fundamentally limiting bandwidths. The first is the response time of the measurement apparatus. It would be defined here by the bandwidth of the resonator, allowing for a time resolution of approximately $20$ ns ($7.8$ MHz). The second limiting factor is the charge relaxation time $T1$ towards a thermal state, since the measurement of the Boltzmann temperature is only meaningful at thermal equilibrium. This happens in this system in about 15 ns (see Supp. Mat. for details)\cite{SuppMat}, whereas maximal relaxation time of about 100 ns have been reported in similar devices\cite{urdampilleta_charge_2015}. DQD thermometry is therefore robust up to a few MHz.

\begin{figure}
    \centering
    \includegraphics[width=\linewidth]{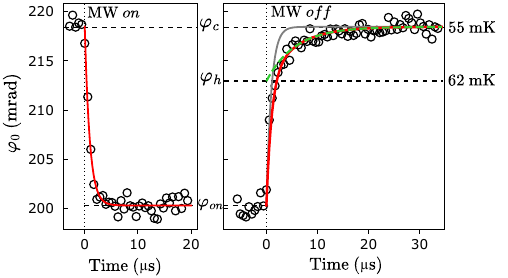}\label{fig:fig4}
    \caption{\textbf{Real-time thermometry with setting II.} Microwave bursts are periodically applied to gate G5 adjacent to the double-dot thermometer. The microwave of frequency 15 GHz and power -40 dBm is modulated by a square signal of frequency 20 kHz, resulting in 50 $\mu$s-long bursts, with 50 $\mu$s waiting time between each burst. The phase signal is acquired with a synchronized lock-in at a sampling rate of $1.7$ MHz with a $400$-ns integration time. It is averaged over $2.10^6$ cycles and is plotted as a function of time (open circles).
    Left panel (heating) : At each cycle, the microwave burst is turned on at $Time=0$, and the phase $\varphi_0$ decreases from $\varphi_c \sim 220$ to $\varphi_{on} \sim 200$ mrad. Right panel (cooling): the microwave excitation is switched off at $Time=0$ and the phase raises back to $\varphi_{on}$ with a two stage relaxation, see plain text for details.}
\end{figure}

For a proof-of-concept demonstration of time-resolved thermometry, we periodically induce local heating by means of microwave bursts applied to a nearby gate, other than the gates defining the double-dot thermometer. Data acquisition is synchronized with the applied bursts to monitor the locally induced temperature modulation \cite{gasparinetti_fast_2015}. 

We plot the resulting phase signals versus time in \hyperref[fig:fig4]{Fig.4}.
We observe the system switching between two phase levels $ \varphi_{on}$ if the microwave is $on$ and $\varphi_{c}$ when it is $off$. This values would translate to $75$ mK and $55$ mK respectively with the previous calibration. However when the drive is $on$, non-adiabatic (Landau-Zener) processes can occur and populate the excited state of the double dot. This would decrease the phase measured. The bath temperature of the "warm" stage can then hardly be precisely known. After the drive is turned $off$, the phase increases back to thermal equilibrium.

The transient regimes exhibit however different characteristic time scales for cooling and heating. 

We model the time evolution after the microwave is turned \textit{on} by an exponential decay:
\begin{equation}
    \varphi_0(\tau) =  \varphi_{on} + (\varphi_{c} - \varphi_{on})\exp \left ( -\tau/\tau_m\right ).
\end{equation}
Which we plot as a red solid lines in \hyperref[fig:fig4]{Fig.4}. We find a characteristic time $\tau_m= 0.93 \pm 0.35$ µs close to our experimental time resolution, which is set by the lock-in measurement bandwidth.

The cooling dynamic reveals two relaxation processes, as evidenced in right panel of \hyperref[fig:fig4]{Fig.4}. The first relaxation corresponds to the rapid thermalization of the the QD thermometer. This occurs on the characteristic time scale for charge relaxation, T1, which we estimate to be about 15~ns (see Supp. Mat. for details)\cite{SuppMat}.  Since $T1 \ll \tau_m$, this charge-relaxation dynamics cannot be resolved and the phase decay is again limited by the measurement bandwidth. A single relaxation on time scale $\tau_m$, however, cannot capture the whole dynamics, as evidenced by the comparison to the grey line in \hyperref[fig:fig4]{Fig.4}.  A second, slower relaxation,  with a characteristic decay time of $\tau_b= 6.0 \pm1.5$ µs, is in fact observed. We ascribe this to the relaxation of the heat bath, which can be revealed once the QD thermometer has thermalized with the surrounding bath. We then propose a more complete model, including a relaxation of the thermometer toward equilibrium with the bath at time $\tau_m$,
\begin{equation}
    \varphi_0(\tau) =  \varphi_{bath}(\tau) + (\varphi_{on} - \varphi_{bath}(\tau))\exp \left ( -\tau/\tau_m\right ),\label{eq:fastrelax}
\end{equation} 
while the bath relaxes at time $\tau_b$ 
\begin{equation}
    \varphi_{bath}(\tau) =  \varphi_{c} + (\varphi_{h} - \varphi_{c})\exp \left ( -\tau/\tau_b\right ).\label{eq:bath}
\end{equation}
Injecting (\ref{eq:bath}) in (\ref{eq:fastrelax}) leads to the red model in the right panel of \hyperref[fig:fig4]{Fig.4} while $\varphi_{bath}$ is plotted with a green dashed line. 
We can extrapolate the relaxation of the bath at $Time = 0 $ to estimate the bath temperature $\sim 62$ mK in the presence of heating, effectively smaller than $75$.  

A deeper investigation (e.g. exploring the effect of varying the microwave excitation cycles and the way they are applied) could provide important missing information about the dissipation mechanisms and the nature of the thermal bath in the QD environment.

Previous works on different types of Si-based spin-qubit devices have indirectly inferred thermal relaxation times from time-resolved Larmor frequency shifts attributed to local overheating \cite{philips_universal_2022-1,takeda_optimized_2018,undseth_hotter_2023} . As compared to this recent literature, the relaxation times reported here are either comparable \cite{takeda_optimized_2018} or significantly shorter \cite{philips_universal_2022-1,undseth_hotter_2023}.  
The discrepancy could be ascribed to the different device structures and to the different experimental conditions (e.g. power and duration of the microwave excitation pulses).

\section{Conclusions}
In summary we have implemented and characterised non-invasive, non-galvanic thermometers sensing either the electronic temperature of a Fermi reservoir, or the local bosonic temperature in a semiconductor quantum-dot device. For the DQD thermometer we could obtain a state-of-the-art noise equivalent temperature of $3$ $\rm mK/\sqrt{\rm Hz}$ and identify a path to even lower noise levels, well below 1 $\rm mK/\sqrt{\rm Hz}$. By synchronizing temperature acquisition to a periodic sequence of microwave bursts, we could increase the signal-to-noise ratio and probe temperature variations on a micro-second time scale.
While this time-resolved QD thermometry is demonstrated in a silicon MOS device, it could be readily  reproduced in any other semiconductor platform, such as SiGe-based heterostructures. 
Hence the present work opens new experimental prospects to understand heating effects in semiconductor quantum processors and to tackle the widely unexplored field of experimental quantum thermodynamics in nanoelectronic systems \cite{pekola_towards_2015}.

\section{Acknowledgements}
This research has been supported by the European Union’s Horizon 2020 research, innovation program under grant agreements number 951852 (QLSI project) and number 810504 (ERC project QuCube), the Program QuantForm-UGA n° ANR-21-CMAQ-0003 France 2030 and by the LabEx LANEF n° ANR-10-LABX-51-01.

\end{document}


\title{
Real-time milli-Kelvin thermometry in a semiconductor qubit architecture
\newline
(Supplemental Material)
}

\author{V. Champain}
\author{V. Schmitt}
\affiliation{Univ. Grenoble Alpes, CEA, Grenoble INP, IRIG-Pheliqs, Grenoble, France.}
\author{B. Bertrand}
\author{H. Niebojewski}
\affiliation{Univ. Grenoble Alpes, CEA, LETI, Minatec Campus, Grenoble, France.}
\author{R. Maurand}
\author{X. Jehl}
\author{C. B. Winkelmann}
\author{S. De Franceschi}
\author{B. Brun}
\affiliation{Univ. Grenoble Alpes, CEA, Grenoble INP, IRIG-Pheliqs, Grenoble, France.}

\date{\today}

\maketitle

\appendix

\section{I. Resonator}
\label{subsec:resonator}
\begin{figure}[h]
    \centering
    \includegraphics[width=\linewidth]{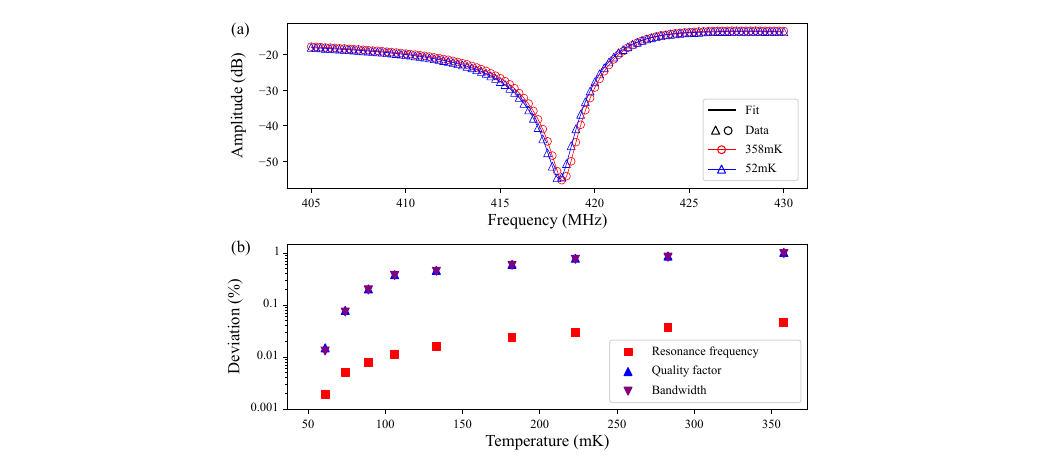}
    \caption{\textbf{Resonator response and its temperature dependence.}\textbf{(a)}Frequency dependence of the reflectometry amplitude for the lowest (52 mK) and highest temperatures (358 mK) in this work (open triangles and circles, respectively). The two data sets are fitted to Eq. (\ref{eq:hanger}) (solid lines) . The resonance frequency shift is much smaller than the resonator linewidth. \textbf{ (b)} Temperature-induced relative deviation of the resonance frequency, quality factor and bandwith of the resonator.}
    \label{fig:fig_supp_1}
\end{figure}
For each measurement temperature, the resonator spectral response is measured to verify that variations measured on the Coulomb peak are not due to some changes in the resonator itself. The spectrum remains essentially unchanged as shown in Fig. S1(a). To extract the resonator characteristics and quantify their temperature-induced deviation, the reflectometry amplitude is fitted to the following equation: 
\begin{equation}
    S_{21, \rm dB} = 20 \log \left ( \left |1+\frac{Q_ie^{i\phi}}{Q_c\left ( 1+ 2iQ_i\left ( \frac{f-f_r}{f_r}\right )\right ) } \right | \right ) +S_0  \label{eq:hanger}
\end{equation}
where $S_{21, \rm dB}$ is the amplitude expressed in dB, $Q_i$ is the internal quality factor, $Q_c$ is the external quality factor, $f_r$ is the resonance frequency, and $S_0$ is the background. The total quality factor $Q$ is given by : 
\begin{equation}
    Q = \left ( \frac{1}{Q_i} + \frac{1}{Q_c}\right )^{-1}
\end{equation}

At 52 mK, we find $Q_i = 385$, $Q_c = 62.0$ and $Q = 53.4$. Its temperature-dependent deviation, $\delta Q$,  is defined as : 
\begin{equation}
    \delta Q = \frac{Q(T) - Q(52\text{ mK})}{Q(52\text{ mK})}
\end{equation}
and analogous definitions hold for the deviations in resonance frequency and linewidth. 
All deviations are found to be smaller than $1\%$ , confirming that the temperature dependence of the measured reflectometry signals cannot be due to undesirable variations in the resonator characteristics.

Note that the time-resolution limit imposed by the resonator is : 
\begin{equation}
    \tau = \frac{Q}{2\pi f_0}
\end{equation}

\section{II. Dot - Lead subsystem}
\label{subsec:dotlead}
\begin{figure}[h]
    \centering
    \includegraphics{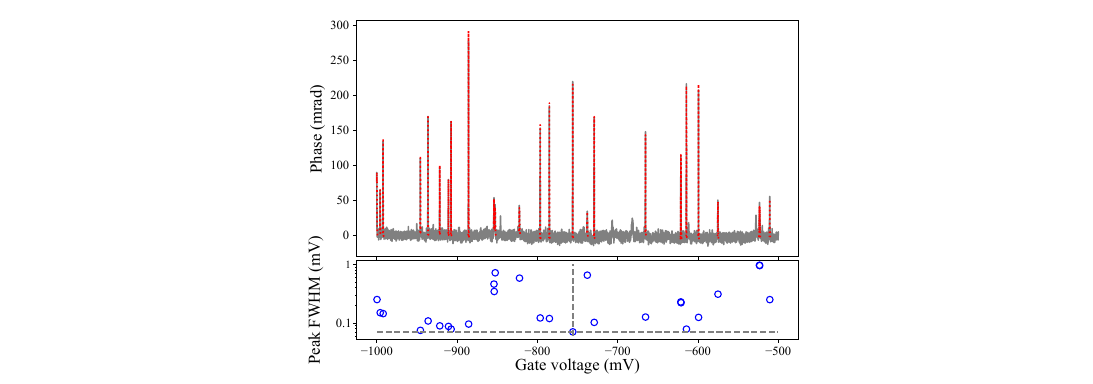}
    \caption{\textbf{Dot-lead transitions.} \textbf{Upper graph:} Reflected phase as a function of gate voltage. Coulomb peaks are measured, corresponding to charging events of the dot. Peaks are detected with a basic threshold detection, and are then fitted to a lorentzian. \textbf{Lower graph:} FWHM of each peak extracted from the fits.}
    \label{fig:dotlead}
\end{figure}

As described in the main text, we chose the dot-lead transition with lowest tunnel coupling. To do that, we sweep the gate voltage of the dot in the few-hole regime from $-500$ to $-1000$ mV, we measure all Coulomb peaks with high gate resolution (Fig. S2(a)) and fit them to a Lorentzian function to extract their full width at half maximum (FWHM), plotted in Fig. S2(b). For the operation of the QD thermometer, we then sweep the gate voltage around the narrowest peak (corresponding to the grey dashed lines in Fig. S2(b)) which is $70$ $\mu$V wide here. 

High RF read-out power can widen the measure dot-lead peaks. Power is then decreased until the width of the narrowest peak becomes as small as possible and power-independent. Integration time is increased accordingly to ensure a sufficiently large signal-to-noise ratio.

Once the QD thermometer is tuned to the narrowest peak, its operation is tested by varying the mixing-chamber temperature through a resistive heater. For each temperature, the peak lineshape is measured by 
averaging the signals from ten identical gate-voltage scans across the peak itself.  

To extract the electronic temperature, $T_e$,  we use the model given by Eq. 1 in the main text, which is the same as in  Ref \cite{chawner_nongalvanic_2021}. From this model we numerically derive a fitting function (for a peak centered around $\varepsilon=0$): 
\begin{equation}
    \varphi(\varepsilon,T_e,\Gamma,A,\alpha) = A\times \alpha^2 e^2\int_{-\infty}^{+\infty }  \left (  \frac{1}{4k_BT_e} \cosh\left (\frac{e \alpha \varepsilon -E}{2k_BT_e}\right )^{-2}\right )\left ( \frac{\hbar \Gamma}{\hbar^2\Gamma^2 + E^2}\right )\text{d}E
\end{equation}

To  numerically calculate the integral, we consider an integration window equal to $10\times \max \left ( 2k_BT_e, h\Gamma \right)$, relying on the peak shape of both functions that vanish out of this window. All peaks are fitted to this formula while imposing a same value of  $\Gamma$, $\alpha$ and $A$. The lever arm $\alpha$ is pre-determined by fitting the high temperature peaks ($T_{MC} >100$ mK) where we assume that $T_e = T_{MC}$ and \begin{equation}
    \varphi(\varepsilon,T_e,\Gamma,\alpha) = A\times \frac{1}{4k_BT_e} \cosh\left (\frac{e \alpha \varepsilon}{2k_BT_e}\right )^{-2}
\end{equation}

\section{III. Isolated double-dot subsystem}
\label{subsec:dqd}

\begin{figure}[h]
    \centering
    \includegraphics[width = \linewidth]{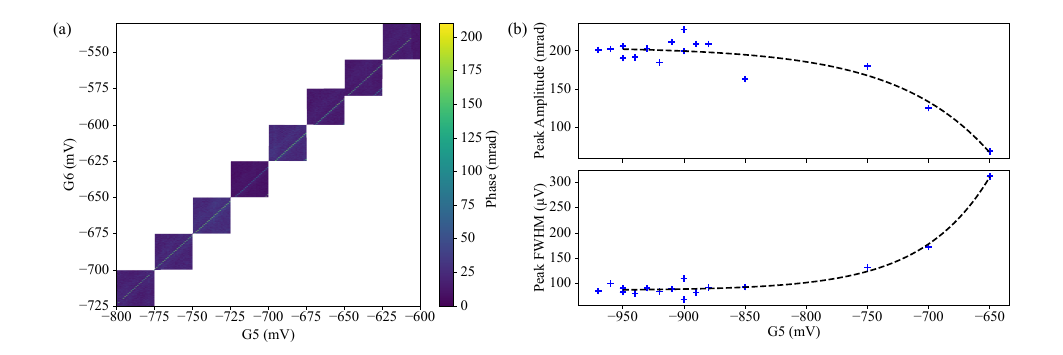}
    \caption{\textbf{Isolated double dot charge stability diagram}. \textbf{(a)} Charge stability diagram showing 'infinite' interdot transitions. \textbf{(b)} Peak amplitude and width as a function of the gate voltage, showing a way to reduce the tunnel coupling.}
    \label{fig:stabilitydoubledot}
\end{figure}

As described in the main text, we first load a few holes in a double QD by applying a negative voltage to G5, G6 and G7, and the bringing back G7 to zero voltage, therefore separating a few charges from the drain reservoir. Provided the inter-dot tunnel coupling is greater that the resonator frequency, inter-dot tunneling transitions can be observed in the reflectometry signal \cite{eenink_tunable_2019,bertrand_quantum_2015}. Since the charges in the double QD are isolated from the leads, the interdot transitions are in fact the only visible features and they extend all over the stability diagram (see \hyperref[fig:stabilitydoubledot]{Fig. S3(a)}). Following the interdot lines towards more negative gate voltages corresponds to making deeper and deeper confinement quantum wells, and simultaneously increasing the height of the inter-dot tunnel barrier,
therefore lowering the tunnel coupling $t$. This trend is experimentally confirmed by recording the interdot transition for different gate voltages. The interdot is simply fitted to a Lorentzian to extract its amplitude and FWHM. Results are shown in \hyperref[fig:stabilitydoubledot]{Fig. S3(b)}. When the voltage of G5 is varied from $-650$ to $-900$ mV, the amplitude increases from $100$ to $200$ mrad and the width simultaneously decreases from $300$ to less than $100$ $\mu$V, indicating that the tunnel coupling is reduced. No quantitative statement can be done since other parameters may slightly change with the electrostatic configuration, such as the gate lever-arm or the charge-photon coupling. 

To express the cross-section of the interdot, we model the double-dot by a charge two-level system, described by the Hamiltonian:
\begin{equation}
    H = \begin{pmatrix}
    \varepsilon/2 & t \\
    t & -\varepsilon/2
    \end{pmatrix}
\end{equation}
Diagonalising this Hamiltonian leads to two eigenstates of energy :
\begin{equation}
    E_{|\pm\rangle} = \mp \sqrt{\left (\frac{\varepsilon}{2}\right ) ^2 + t^2 }
\end{equation}
As represented in Fig.1(e), we have then two available anticrossing states with an energy gap  $\Delta E = \sqrt{ \varepsilon ^2 + (2t)^2 }$ in contact with a local environment at temperature $T_{ph}$. Each state is thermally populated according to a Boltzmann distribution  : 
\begin{align}
    P_{|\pm\rangle} &= \frac{e^{\frac{\pm\Delta E}{2k_BT}}}{Z}\\
    Z &= e^{\frac{\Delta E}{2k_BT}} + e^{\frac{-\Delta E}{2k_BT}}\\
\end{align}

Each state results in a quantum-capacitance contribution \cite{penfold-fitch_microwave_2017,mizuta_quantum_2017,urdampilleta_charge_2015,petersson_charge_2010} : 
\begin{equation}
    C_q^{|\pm\rangle} = \frac{\partial^2  E_{|\pm\rangle}}{\partial V^2 }
\end{equation}
Where $V = \varepsilon/\alpha e $. We can then express the total quantum capacitance as: 
\begin{equation}
    C_q =  C_q^{|+\rangle} P_{|+\rangle} +C_q^{|-\rangle} P_{|-\rangle} =  \alpha^2 e^2\frac{2t^2}{(\varepsilon^2+4t^2)^{3/2}}\tanh\left(  \frac{\left ( \varepsilon^2+4t^2\right )^{1/2}}{2k_BT}\right)    
\end{equation}
From this equation we define the fitting function : 
\begin{equation}
    \varphi(V,T, t,A,\alpha) = A\times \frac{2t^2}{((\alpha eV)^2+4t^2)^{3/2}}\tanh\left(  \frac{\left ( (\alpha e V)^2+4t^2\right )^{1/2}}{2k_BT}\right)  
\end{equation}

The temperature of the fridge is varied by means of resistive heater mounted on the mixing chamber. For each temperature, the selected peak is measured ten times for statistics. As explained in the main text, the peak amplitude is highly temperature dependent containing all the needed information. As a result, we fix the $t$ and $A$ parameters and leave the peak amplitude as the only temperature-dependent fitting parameter. With theses values, the peaks are then individually fitted, leaving only $T$ and $\alpha$ as free parameter. These parameters are independent, and can be fitted without ambiguity as the temperature impacts the amplitude of the peak while the lever arm affects the width. For each bath temperature, we average the temperatures obtained from the fits, and take the standard deviation . The results are reported in Fig. 2(d).

The dispersive shift of the resonator at resonance can be expressed as\cite{blais_circuit_2021}
\begin{equation}
    \chi = g_0^2\left (\frac{1}{2t+f_r}+\frac{1}{2t-f_r} \right ) 
\end{equation}
where $g_0$ is the bare charge photon coupling and $f_r$ is the resonance frequency of the resonator, which we use to compute the scaling of the parameter $\kappa$ in Eq. (3).

\section{IV. Noise equivalent temperature}

\begin{figure}[h]
    \centering
    \includegraphics[width = \linewidth]{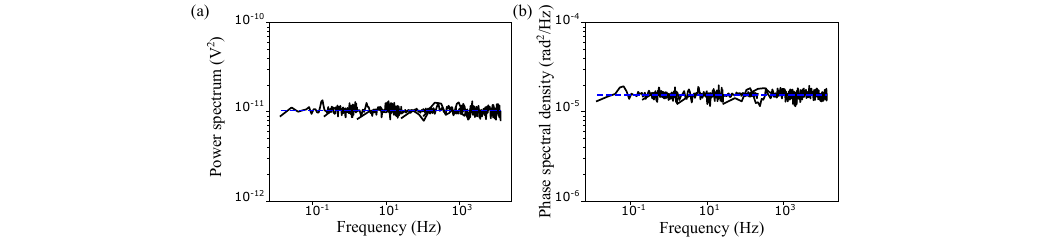}
    \caption{\textbf{Noise spectrum.} \textbf{a)} Power spectrum computed using Welchs'method\cite{welch_use_1967}, showing a white noise spectrum with a noise floor $S_P = 1.03 \pm 0.09 \times 10^{-11}\rm V^2 $.  \textbf{b)} Phase spectral density, showing a white noise spectrum with a noise floor $S_{\varphi \varphi} = 3.9 \pm 0.2\text{ mrad}/\sqrt{\text{Hz}}$}
    \label{fig:NET}
\end{figure}
\label{subsec:noise}
The two quadratures of the demodulator $I$ and $Q$ are measured for different sampling frequencies to access  noise spectrum in a wide frequency range.  From the noise floor we can access to the noise temperature $NT$ : 
\begin{equation}
    NT = \frac{S_P}{10 ^\frac{G_{dB}}{10}B\times 4k_BR}
\end{equation}
Where $S_P$ is the noise floor, $G_{dB}$ is the gain in dB of the room temperature amplifier, $B$ is the bandwidth of the demodulator and $R$ is the $50\ohm$-impedance of the transmission line. We find a $NT = 1.9\pm0.3$ K, where as the cryogenic amplifier (LNF-LNC 0.2-3 A s/n 1410Z) used in this experiment has a noise temperature of around $2$ K 

\section{V. Evidence of different relaxation time scales}

\begin{figure}[h]
    \centering
    \includegraphics[width = \linewidth]{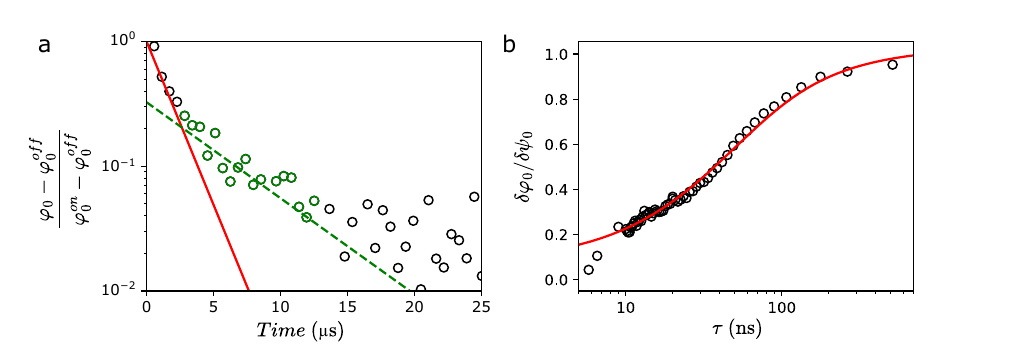}
    \caption{\textbf{Time-resolved thermometry and charge relaxation time. a.}  Data of Fig.4 in the plain text, plotted as a relative temperature change in log scale versus time. Two different relaxation time scales are evidenced by two linear fits to the data, red and green. \textbf{b.} Charge relaxation time T1 using the method of chopped microwaves. The red curve shows a fit giving T1 = 14 ns }
    \label{fig:relaxation}
\end{figure}
The data for the cooling dynamics are rescaled and plotted in a logarithmic axis to highlight two distinct relaxations mechanisms (Fig. S5a). The first one, highlighted by the red line, corresponds to a fast relaxation, i.e. the exponential decay of Fig.4. The second one, highlighted by the green dashed line, points to a second, slower relaxation mechanism, corresponding to the green-line exponential decay in Fig. 4. As the thermometer used in this experiment is equivalent to a charge qubit, we can only consider the system being in equilibrium with the thermal bath at times longer than $T_1$. We evaluate this relaxation time with the 'chopped microwaves' method described in \cite{petta_manipulation_2004}. A microwave of frequency $2t$, resonant with the charge qubit, drives the system to an incoherent mixture of ground and excited state. We modulate the microwave by a square envelope with a 50\% duty-cycle. If the period is smaller than the relaxation time, the system stays in average in the mixed state. But if the period is longer than $T_1$, then the system has time to relax. In average, the system will be 50\% of the time in the mixed state, and 50\% of the time in the ground state. By recording the averaged phase as a function of the cycle period, we can fit the value of $T_1$ with the model \begin{equation}
    \delta\varphi_0/ \delta\psi_0 = \frac{\varphi(\tau)-\varphi(\tau = 0)}{\varphi(\tau = +\infty)-\varphi(\tau = 0)}= 1-\frac{2T_1}{\tau} \left( 1-\exp \left (-\frac{\tau}{2T_1} \right ) \right )
\end{equation}
We extract $T_1 \sim 15$ ns. The first relaxation measured corresponds then to the bandwidth of the measurement. The second relaxation must be of thermal origin.
\section{VI. Measured devices}
Three different devices with the same geometry have been measured, i.e. a nanowire cross section of 80 $\times$16 nm$^2$ and a gate pitch of 80 nm. Results are reproducible from one device to another. 
\begin{table}[h]
    \centering
    \begin{tabular}{c|c}
        Sample & Figure \\ \hline
        1 & Fig. 2.(a) and (c) \\
        2 & Fig. 2.(b) and (d) \\
        1 & Fig. 3. \\
        3 & Fig. 4.
    \end{tabular}
    \caption{Device associated to each dataset presented in the plain text}
\end{table}

\section{VII. DC filters}
\begin{figure}[h]
    \centering
    \includegraphics{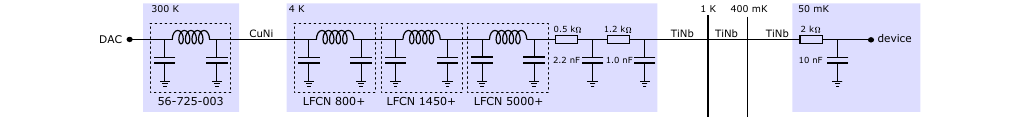}
    \caption{Schematic of a DC-line with its corresponding filters. Filters cut-off frequencies are respectively from left to right : 3 MHz, 80 MHz, 1.5 GHz, 5 GHz, 50 kHz, 50 kHz, 2 kHz. }
    \label{fig:dcfiltering}
\end{figure}
Fig \ref{fig:dcfiltering}. shows the DC-filtering in this experiment, as high-frequency filtering is of primary importance to decrease the electronic temperature. Wires are thermalized at each temperature stage by copper stripes. At the mixing-chamber stage, the loom is sunk in conductive epoxy to imitate a copper powder filter.

Achieving a lower electronic temperature can be challenging. Recent works \cite{nicoli_quantum_2019} used a He$^4$ immersion cell to decrease their electronic temperature up to 6.7 mK. They  also noticed that mechanical vibrations can have a detrimental impact on the electronic temperature.

\section{VIII. Schematic of the device}
\begin{figure}[h]
    \centering
    \includegraphics[scale=.7]{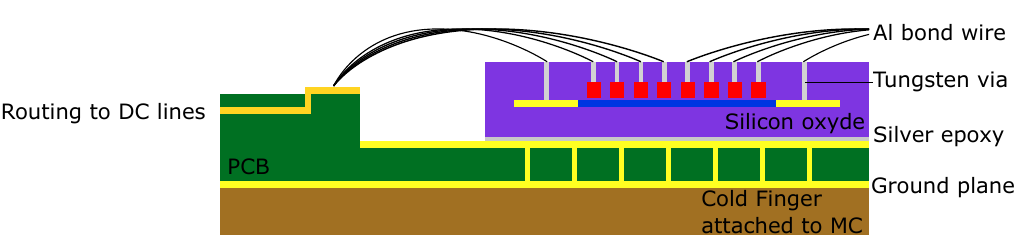}
    \caption{Schematic of the device. }
    \label{fig:dcfiltering}
\end{figure}